\documentclass[12pt,reqno]{amsart}

\usepackage{amsmath,
amsfonts,
amsthm,
cite
}

\usepackage{a4wide}

\usepackage{times,txfonts}

\usepackage{psfrag}
\usepackage[dvips]{graphicx}
\usepackage{epsf,epsfig}

\DeclareMathOperator{\Tr}{Tr}
\DeclareMathOperator{\sh}{sh}
\DeclareMathOperator{\ch}{ch}

\DeclareMathOperator{\CS}{CS}

\usepackage{titlesec}
\titleformat{\section}
{\bfseries\scshape\centering}
{\thesection.}{.5em}{}
\titleformat{\subsection}
{\rmfamily\bfseries}
{\thesubsection}{.5em}{}

\begin{document}
\baselineskip 16.8pt
\parskip 7pt


\title{
  Torus Knot and Minimal Model
}


    \author{Kazuhiro \textsc{Hikami}}


  \address{Department of Physics, Graduate School of Science,
    University of Tokyo,
    Hongo 7--3--1, Bunkyo, Tokyo 113--0033, Japan.
    }

    \email{\texttt{hikami@phys.s.u-tokyo.ac.jp}}
    \author{Anatol N. \textsc{Kirillov}}
    \address{RIMS, Kyoto University,  Kyoto 606-8502, Japan.}
    \email{\texttt{kirillov@kurims.kyoto-u.ac.jp}}

\date{August 22, 2003}

\begin{abstract}
  We reveal an intimate connection between the quantum knot invariant
  for torus knot $T(s,t)$ and the  character of the minimal
  model $\mathcal{M}(s,t)$, where $s$ and $t$ are relatively prime
  integers.
  We show that
  Kashaev's invariant,
  \emph{i.e.},  the $N$-colored Jones polynomial  at the $N$-th root
  of unity,
  coincides with the Eichler integral of the
  character.
\end{abstract}





\maketitle

\section{Introduction}

After Jones polynomial was introduced~\cite{Jones85}, 
studies of  quantum
invariants have been extensively developed.
These quantum knot invariants are physically interpreted
as the Feynman path integral of
the Wilson loop with the
Chern--Simons action~\cite{EWitt89a}.
Though,
geometrical
interpretation of the quantum invariant is still not complete.
Some time ago, Kashaev defined a quantum knot invariant
based on the
quantum dilogarithm function~\cite{Kasha95},
and made a conjecture~\cite{Kasha96b} that a limit of his
invariant coincides with the hyperbolic volume
of the knot complement~\cite{WPThurs80Lecture}.
This suggests an intimate connection between the quantum invariant
and
the geometry.
Note that Kashaev's invariant was later identified with a specialization of the
$N$-colored Jones polynomial at $q$ being the $N$-th primitive root of
unity~\cite{MuraMura99a}.

\begin{figure}[htbp]
  \centering
  
  \includegraphics{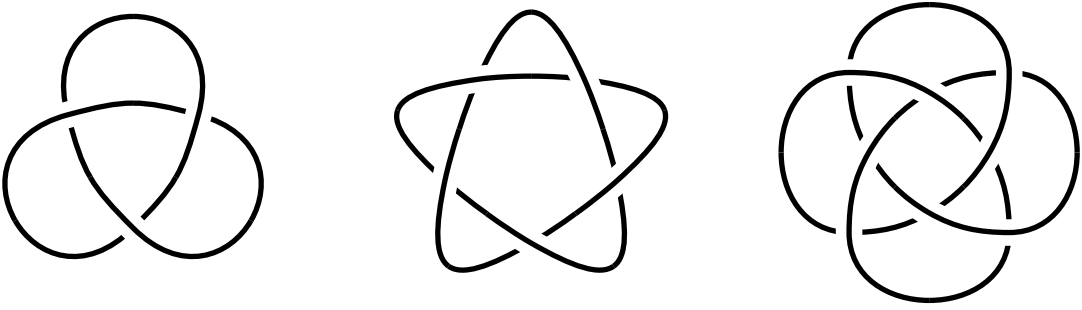}
 
  \caption{Torus knot $T(s,t)$.
    {}From left to right, we depict
    trefoil $T(2,3)$, Solomon's seal knot $T(2,5)$, and
    $T(3,4)$, respectively.
  }
  \label{fig:torus}
  
\end{figure}

In this article, we study Kashaev's invariant
$\langle \mathcal{K} \rangle_N$ for the torus knot
$\mathcal{K}= T(s,t)$, where $s$ and $t$ are coprime.
See Fig.~\ref{fig:torus} for a projection of some torus knots.
One may think that
it is  insignificant from a view point of the \emph{Volume Conjecture}
because
the torus knot is not hyperbolic~\cite{WPThurs80Lecture}.
Although,
the Chern--Simons invariant is considered as an imaginary part of the
hyperbolic volume, and in fact
the torus knot is supposed to have
non-trivial Chern--Simons   invariant.
We shall  show that the invariant 
exactly  coincides with a limiting value of the
Eichler integral of
the character of the minimal model $\mathcal{M}(s,t)$
with $q$ being the $N$-th root of unity.

This paper is organized as follows.
In Section~\ref{sec:character} we recall a modular property of the
character of the minimal model $\mathcal{M}(s,t)$.
We define the Eichler integral, and give an explicit form
of limiting value thereof when $q$ is the $N$-th primitive root of unity.
In Section~\ref{sec_invariant}
we study the colored Jones polynomial for the torus knot $T(s,t)$.
We give a formula relating the quantum invariant with the Eichler
integral.
We further give some examples on $q$-series identities.
Clarified is a relationship between the conformal weight and the
Chern--Simons invariant of the minimal model.
The last section is devoted to concluding remarks.

\section{Eichler Integral of the  Character}
\label{sec:character}

We study the character of the minimal model $\mathcal{M}(s,t)$, where
$s$ and $t$ are coprime integers.
The central charge of the minimal model $\mathcal{M}(s, t)$  is
\begin{equation}
  c(s,t) = 1 - \frac{6 \, (s-t)^2}{s \, t} ,
\end{equation}
and the irreducible highest weight representation of the Virasoro
algebra is given for the conformal weight
\begin{equation}
  \Delta^{s,t}_{n,m}
  =
  \frac{(n \, t - m \, s)^2 - (s-t)^2}{4 \, s \, t} ,
\end{equation}
where integers $m$ and $n$ are
\begin{align*}
  & 1 \leq n \leq s-1,
  &
  & 1 \leq m \leq t- 1.
\end{align*}
The number of distinct fields in the theory is 
\begin{equation}
  D(s,t) = \frac{1}{2} \, (s-1 ) \, (t-1) .
\end{equation}

The character $\ch_{n,m}^{s,t}(\tau)$ for an irreducible highest weight 
representation of the Virasoro algebra with above central charge and
weight,
is computed as~\cite{Rocha84a,BFeigDBFuk83a}
\begin{align}
  \ch_{n,m}^{s,t}(\tau)
  & =
  \Tr  q^{L_0 - \frac{1}{24} c(s,t)}
  \nonumber \\
  & =
  \frac{q^{\Delta^{s,t}_{n,m} - \frac{1}{24} c(s,t)}}{(q)_\infty} \,
  \sum_{k \in \mathbb{Z}}
  q^{s t k^2} \,
  \left(
    q^{k ( n t - m s)} - q^{k ( n t +m s)+ m n}
  \right) ,
\end{align}
where we set $q=\mathrm{e}^{2 \pi \mathrm{i} \tau}$.
We see that
\begin{equation*}
  \ch_{n,m}^{s,t}(\tau)
  =
  \ch_{s-n,t-m}^{s,t}(\tau)
  =
  \ch_{m,n}^{t,s}(\tau)
  =
  \ch_{t-m,s-n}^{t,s}(\tau) .
\end{equation*}
The  character is modular
covariant~\cite{ItzyZube86,CappItzyZube87b} as
\begin{equation}
  \ch_{n,m}^{s,t}(\tau)
  =
  \sum_{n^\prime , m^\prime}
  \mathbf{S}_{n,m}^{n^\prime , m^\prime} \,
  \ch_{n^\prime , m^\prime}^{s,t} (-1 / \tau) ,
\end{equation}
where sum runs over $D(s,t)$ distinct fields, and a matrix is
explicitly written as
\begin{equation}
  \mathbf{S}_{n,m}^{n^\prime , m^\prime}
  =
  \sqrt{\frac{8}{\   s \, t   \  }} \,
  (-1)^{n  m^\prime + m n^\prime +1} \,
  \sin
  \left(
     n \, n^\prime \, \frac{\ t  \  }{s}  \,  \pi 
  \right) \,
  \sin
  \left(
      m \, m^\prime \, \frac{ s}{ \  t  \  } \, \pi
  \right) .
\end{equation}

We rewrite the   character of the minimal model as
\begin{equation}
  \ch_{n,m}^{s,t}(\tau)
  =
  \frac{
    \Phi^{(n,m)}(\tau)
  }{\eta(\tau)} .
\end{equation}
Here
we have set the Dedekind $\eta$-function and $\Phi^{(n,m)}(\tau)$ as
\begin{gather}
  \eta(\tau)
  =
  q^{1/24} \, (q)_\infty, 
  \nonumber
  \\[2mm]
  \Phi^{(n,m)}(\tau)
  =
  \sum_{k=0}^\infty \chi_{2 s t}^{(n,m)}(k) \, q^{\frac{1}{4 s t} k^2} ,
  \label{define_Phi}
\end{gather}
where the function $\chi_{2 s t}^{(n,m)}(k)$ is periodic with modulus
$2\, s \, t$ as
\begin{equation}
  \begin{array}{c|ccccc}
    k
    \mod 2 \, s \, t
    & n \, t - m \, s & n \, t + m \, s &
    2 \, s \, t - ( n \, t + m \, s)
    & 2 \, s \, t - (n \, t - m \, s) &
    \text{others}
    \\
    \hline
    \chi_{2 s t}^{(n,m)}(k)
    &
    1 & -1 & -1 & 1 & 0
  \end{array}
\end{equation}
{}From the modular property of the Dedekind $\eta$-function,
we see that $\Phi^{(n,m)}(\tau)$ is modular with weight $1/2$, and
spans $D(s,t)$ dimensional space;
modular
$T$- and $S$-transformations are respectively written as
\begin{gather}
  \Phi^{(n,m)}(\tau+1)
  =
  \mathrm{e}^{\frac{(n t - m s)^2}{2 s t} \pi \mathrm{i}} \,
  \Phi^{(n,m)}(\tau) ,
  \\[2mm]
  \Phi^{(n,m)}(\tau)
  =
  \sqrt{\frac{\   \mathrm{i}  \  }{\tau}} \,
  \sum_{n^\prime , m^\prime}
  \mathbf{S}_{n,m}^{n^\prime , m^\prime} \,
  \Phi^{(n^\prime  ,   m^\prime)}(-1/\tau) .
  \label{Phi_S_transform}
\end{gather}

For the modular form with weight $w \in \mathbb{Z}_{>2}$, the period
is defined by use of the classical Eichler integral, which is $w-1$
integrations of the modular form with respect to
$\tau$~\cite{SLang76Book}.
In a case of the half-integral weight modular form
$\Phi^{(n,m)}(\tau)$,
the Eichler integral is thus naively defined by the $q$-series as~\cite{DZagie01a}
\begin{equation}
  \label{define_tilde}
  \widetilde{\Phi}^{(n,m)}(\tau)
  =
  -\frac{1}{2} \, \sum_{k=0}^\infty k \, 
  \chi_{2 s t}^{(n,m)}(k) \, q^{\frac{1}{4 s t} k^2} .
\end{equation}
A prefactor is for our convention.
It can be seen that
the former is regarded as a ``half-derivative''
($\frac{1}{2}-1$ integration) of the modular form
$\Phi^{(n,m)}(\tau)$ with respect to $\tau$,
as was originally studied in Ref.~\citen{DZagie01a}.
We consider a limiting value of the Eichler integral
$\widetilde{\Phi}^{(n,m)}(\alpha)$ at $\alpha  \in \mathbb{Q}$.
Applying the Mellin transformation, we have
\begin{equation*}
  \widetilde{\Phi}^{(n,m)}
  \left(
    \frac{M}{N} +
    \mathrm{i} \, \frac{y}{2 \,  \pi}
  \right)
  \simeq
  - \frac{1}{2} \, \sum_{k=0}^\infty
  \frac{L_\omega(- 2 \, k -1 , \chi_{2 s t}^{(n,m)} )}{k!} \,
  \left(
    - \frac{y}{4 \,s \, t}
  \right)^k ,
\end{equation*}
where $y\searrow 0$, and $M,N$ are coprime integers.
We mean that $L_\omega (k,\chi_{2 s t }^{(n,m)})$ is the twisted
$L$-function defined by
\begin{align*}
  L_\omega(k, \chi_{2 s t}^{(n,m)})
  & =
  \sum_{j=1}^\infty \chi_{2 s t}^{(n,m)} (j) \,
  \mathrm{e}^{ \frac{M}{N} \frac{j^2}{2 s t}  \pi \mathrm{i}} \,
  j^{-k}
  \\
  & =
  \frac{1}{(2 \, s \, t \, N)^k} \,
  \sum_{j=1}^{2 s t N}
  \chi_{2 s t}^{(n,m)}(j) \,
  \mathrm{e}^{ \frac{M}{N} \frac{j^2}{2 s t} \pi \mathrm{i}} \,
  \zeta \left(k, \frac{j}{2 \, s \, t \, N}\right)
  ,
\end{align*}
where $\zeta(k,x)$ is the Hurwitz $\zeta$ function.
By the analytic continuation,
limiting value at $\tau \to M/N $ is then computed as
\begin{equation}
  \label{Eichler_for_MN}
  \widetilde{\Phi}^{(n,m)}(M/N)
  =
  \frac{s \, t \, N}{2}
  \sum_{k=1}^{2 s t N}
  \chi_{2 s t}^{(n,m)}(k) \,
  \mathrm{e}^{\frac{k^2 M}{2 s t N} \pi \mathrm{i}} \, 
  B_2\left(\frac{k}{2 \, s \, t \, N}\right) ,
\end{equation}
where $B_k(x)$ is the $k$-th Bernoulli polynomial,
$
\displaystyle
\frac{t \, \mathrm{e}^{x t}}{\mathrm{e}^t - 1}
=
\sum_{k=0}^\infty  \frac{t^k}{k!} \, B_k(x)
$, and especially
$B_2(x) = x^2 - x + \frac{1}{6}$.

This function  fulfills a \emph{nearly} modular property;
for $N \in \mathbb{Z}$ we have an asymptotic expansion in
$N\to \infty$,
\begin{multline}
  \label{nearly_modular_Phi}
  \widetilde{\Phi}^{(n,m)}(1/N)
  + ( - \mathrm{i} \, N)^{3/2} \,
  \sum_{n^\prime , m^\prime}
  \mathbf{S}_{n,m}^{n^\prime , m^\prime} \,
  \phi(n^\prime , m^\prime) \,
  \mathrm{e}^{
    - \frac{(n^\prime t - m^\prime s)^2}{2 s t} \pi \mathrm{i} N
  }
  \\
  \simeq
  \sum_{k=0}^\infty
  \frac{T^{(n,m)}(k)}{k!} \,
  \left(
    \frac{\pi}{2 \, s \, t \, \mathrm{i} \, N}
  \right)^k .
\end{multline}
Here we have set
\begin{equation}
  \phi(n,m)
  =
  \begin{cases}
    (s-n) \, m ,
    &
    \text{if  $n \,t > m \, s$,}
    \\[2mm]
    n \, (t-m) ,
    &
    \text{if  $n \, t < m \, s$,}
  \end{cases}
\end{equation}
and  $T$-series is written in terms of the $L$-function associated
with $\chi_{2 s t}^{(n,m)}$ as
\begin{align}
  T^{(n,m)}(k)
  & =
  \frac{1}{2} \, (-1)^{k+1} \, L(-2 \, k -1 , \chi_{2 s t}^{(n,m)})
  \nonumber \\
  & =
  \frac{1}{2} \, (-1)^k \, \frac{(2 \, s \, t)^{2 k+1}}{2 \, k +2} \,
  \sum_{j=1}^{2 s t}
  \chi_{2 s t}^{(n,m)}(j) \,
  B_{2 k +2} \left( \frac{j}{2 \, s \, t} \right) .
\end{align}
This can be
shown as follows
(see Refs.~\citen{DZagie01a,LawrZagi99a,KHikami02c,KHikami03a}).
We define a variant of the Eichler integral
\begin{equation}
  \label{variant_Eichler}
  \widehat{\Phi}^{(n,m)}(z)
  =
  \sqrt{\frac{s \, t \, \mathrm{i}}{8 \, \pi^2}}
  \,
  \int_{z^*}^\infty
  \frac{
    \Phi^{(n,m)}(\tau)
  }{
    (\tau - z)^{3/2}
  } \,
  \mathrm{d} \tau .
\end{equation}
This
function is defined for $z$ in the lower half plane,
$z \in \mathbb{H}^-$,
while  the Eichler integral $\widetilde{\Phi}^{(n,m)}(z)$ is for the
upper half plane, $z \in \mathbb{H}$.
Using $S$-transformation~\eqref{Phi_S_transform}, we have
\begin{equation}
  \label{Eichler_S_transform}
  \widehat{\Phi}^{(n,m)}(z)
  +
  \left(
    \frac{1}{\mathrm{i} \, z}
  \right)^{3/2} \,
  \sum_{n^\prime , m^\prime} \mathbf{S}_{n,m}^{n^\prime , m^\prime} \,
  \widehat{\Phi}^{(n^\prime , m^\prime)}(-1/z)
  =
  r^{(n,m)}(z; 0),
\end{equation}
where we have defined the period function
\begin{equation}
  r^{(n,m)}(z; \alpha )
  =
  \sqrt{\frac{s \, t \, \mathrm{i}}{8 \, \pi^2}}
  \,
  \int_{\alpha}^\infty
  \frac{
    \Phi^{(n,m)}(\tau)
  }{
    (\tau - z)^{3/2}
  } \,
  \mathrm{d} \tau ,
\end{equation}
for $z \in \mathbb{H}^-$ and $\alpha \in \mathbb{Q}$.
More generally, for
$\gamma = 
\begin{pmatrix}
  a & b \\
  c & d
\end{pmatrix}
\in
SL(2; \mathbb{Z})$,
we have
\begin{equation}
  \widehat{\Phi}^{(n,m)}(z)
  -
  \frac{1}{v^{(n,m)}(\gamma)} \, (c \, z+ d)^{-3/2} \,
  \sum_{n^\prime , m^\prime}
  \left(
    \mathbf{M}_\gamma
  \right)_{n,m}^{n^\prime , m^\prime} \,
  \widehat{\Phi}^{(n^\prime , m^\prime)}  ( \gamma(z) )
  =
  r^{(n,m)}(z ; \gamma^{-1}(\infty)) ,
\end{equation}
where  a matrix $\mathbf{M}_\gamma$ and $v^{(n,m)}(\gamma)$ are given from the
modular transformation,
\begin{equation*}
  \sum_{n^\prime , m^\prime}
  \left(
    \mathbf{M}_\gamma
  \right)_{n, m}^{n^\prime ,m^\prime}
  \,
  \Phi^{(n^\prime, m^\prime)}(\gamma(z))
  =
  v^{(n, m)}(\gamma) \,
  \sqrt{c \, z + d} \,
  \Phi^{(n,m)}(z) .
\end{equation*}
When we substitute eq.~\eqref{define_Phi}
into eq.~\eqref{variant_Eichler} and perform an integration term by
term in a limit  $z \to \alpha  \in \mathbb{Q}$,
we see that
\begin{equation*}
  \widetilde{\Phi}^{(n,m)}(\alpha)
  =
  \widehat{\Phi}^{(n,m)}(\alpha) ,
\end{equation*}
Note that the left hand side is given by eq.~\eqref{Eichler_for_MN} as
a limit value from $\mathbb{H}$ while the right hand side is a limit
from $\mathbb{H}^-$.
We can check for $N\in \mathbb{Z}$ that
an asymptotic expansion of $r^{(n,m)}(1/N;0)$  gives a right hand side
of eq.~\eqref{nearly_modular_Phi}, and that from
eq.~\eqref{Eichler_for_MN}
we have
\begin{gather*}
  \widetilde{\Phi}^{(n,m)}(N+1)
  =
  \mathrm{e}^{\frac{(n t - m s)^2}{2 s t} \pi \mathrm{i}} \,
  \widetilde{\Phi}^{(n,m)}(N),
  \\[2mm]
  \widetilde{\Phi}^{(n,m)}(0)
  =
  \phi(n,m) ,
\end{gather*}
which shows
\begin{equation*}
  \widetilde{\Phi}^{(n,m)}(N)
  =
  \phi(n,m) \,
  \mathrm{e}^{\frac{(n t - m s)^2}{2 s t} \pi \mathrm{i} N} .
\end{equation*}
Combining these results
we recover eq.~\eqref{nearly_modular_Phi}.

\section{Quantum Knot Invariant for Torus Knot}
\label{sec_invariant}

We study  the $N$-colored Jones polynomial
$J_N(\mathcal{K})$ for the torus knot
$\mathcal{K}=T(s,t)$.
The torus knot $T(s,t)$ for coprime integers $s, t$
is the knot which wraps around the solid torus in the longitudinal
direction $s$ times and in the meridinal direction $t$ times.
See Fig.~\ref{fig:torus}.
It
is also  represented as
$( \sigma_1 \, \sigma_2 \cdots \sigma_{s-1} )^t$ in terms of generators $\sigma_j$ of
the Artin braid group.
An explicit form of the $N$-colored Jones polynomial
is read as~\cite{Mort95a,RossJone93a}
\begin{equation}
  \label{colored_polynomial}
  2 \sh (N \, \hbar /2) \,
  \frac{J_N(\mathcal{K})}{J_N(\mathcal{O})}
  = \mathrm{e}^{- \frac{\hbar}{4} (\frac{t}{s} + \frac{s}{t})} \,
  \sum_{\varepsilon = \pm 1}
  \sum_{k=-\frac{N-1}{2}}^{\frac{N-1}{2}}
  \varepsilon \,
  \exp
  \left(
    \hbar \, s \, t \,
    \left(
      k + \frac{ s + \varepsilon \, t}{2 \, s \, t}
    \right)^2
  \right) ,
\end{equation}
where we have set a parameter $q=\mathrm{e}^\hbar$, and $\mathcal{O}$
denotes unknot.
As was shown in Ref.~\citen{MuraMura99a},
Kashaev's invariant~\cite{Kasha95,Kasha96b} coincides with
a specialization $q \to \mathrm{e}^{2 \pi \mathrm{i}/N}$ of the colored
Jones polynomial.
As the left hand side of eq.~\eqref{colored_polynomial} vanishes in
this substitution, Kashaev's invariant for the torus knot
can be computed as
a derivative of the right hand side with respect to $\hbar$.

Here we recall  the Eichler integral
$\widetilde{\Phi}^{(n,m)}(1/N)$
which was computed in
eq.~\eqref{Eichler_for_MN},
and especially pay attention to 
a case of $(n,m)=(s-1, 1)$.
Using a property of the Gauss sum,
we  obtain  from eq.~\eqref{Eichler_for_MN}
\begin{equation}
  \widetilde{\Phi}^{(s-1,1)}(1/N)
  =
  \frac{s \, t}{N} \,
  \mathrm{e}^{
    \frac{s t}{2} N \pi \mathrm{i} + (s+t)
    \pi \mathrm{i}
  }
  \sum_{\varepsilon=\pm 1}
  \sum_{k=-\frac{N-1}{2}}^{\frac{N-1}{2}}
  \varepsilon \,
  \left(
    k+ \frac{s + \varepsilon \, t}{2 \, s \, t}
  \right)^2 \,
  \mathrm{e}^{
    \frac{2 \pi \mathrm{i}}{N} s t
    \left( k + \frac{s + \varepsilon t}{2 s t} \right)^2
  } .
\end{equation}
As seen from eq.~\eqref{colored_polynomial}, this expression is
proportional to 
the colored Jones polynomial at 
$\hbar \to 2 \, \pi \, \mathrm{i}/N$.
To conclude
Kashaev's invariant
$\langle \mathcal{K} \rangle_N$
for torus knot $\mathcal{K}=T(s,t)$
is identified with
\begin{equation}
  \label{invariant_character}
  \mathrm{e}^{-\frac{(s t - s - t)^2}{2 s t N} \pi \mathrm{i}}
  \cdot
  \widetilde{\Phi}^{(s-1,1)}(1/N)
  =
  \langle T(s,t) \rangle_N .
\end{equation}
We expect that
the Eichler integrals $\widetilde{\Phi}^{(n,m)}(1/N)$ for
other cases $(n,m)$ are related with the quantum
invariants of 3-manifolds.
As a result
eq.~\eqref{nearly_modular_Phi} denotes an asymptotic expansion of
Kashaev's invariant in $N\to\infty$.
Note that an asymptotic behavior  was studied in
Refs.~\citen{KHikami02b,KashaTirkk99a} in a different manner.

In general,
we can construct
$q$-series  for the Eichler integrals
based on the $R$-matrix~\cite{Kasha95}.
We give some examples below
(see Fig.~\ref{fig:torus}).
Hereafter we use a standard notation,
\begin{gather*}
  (x)_k = (x ; q)_k = \prod_{j=1}^k (1- x \, q^{j-1}) ,
  \\[2mm]
  \begin{bmatrix}
    \  k  \  \\
    j
  \end{bmatrix}
  =
  \frac{ (q)_k}{ (q)_j \, (q)_{k-j} } .
\end{gather*}
\begin{itemize}
\item Trefoil $T(2,3)$,
  \begin{align*}
    \widetilde{\Phi}^{(1,1)}(\tau)
    & \equiv
    -\frac{1}{2} \, \sum_{k=0}^\infty k \, 
    \chi_{12}^{(1,1)}(k) \, q^{ k^2 / 24} 
    \\
    & =
    q^{1/24} \,
    \sum_{k=0}^\infty (q)_k .
  \end{align*}
  This  equality  is
  Zagier's ``strange'' identity~\cite{DZagie01a};
  though both expressions do not converge simultaneously,
  the limiting values in $q$ being roots of unity coincide.
  It  is  the Eichler integral of the Dedekind
  $\eta$-function.

  \item Solomon's Seal knot
    $T(2,5)$,
    \begin{align*}
      \widetilde{\Phi}^{(1,1)}(\tau)
      & \equiv
      -\frac{1}{2} \, \sum_{k=0}^\infty k \, 
      \chi_{20}^{(1,1)}(k) \, q^{ k^2 / 40} 
      \\
      & =
      q^{9/40} \,
      \sum_{k=0}^\infty (q)_k \,
      \sum_{j=0}^k q^{j (j+1)} \,
      \begin{bmatrix}
        \  k  \  \\
        j
      \end{bmatrix} ,
      \displaybreak[0]
      \\[2mm]
      \widetilde{\Phi}^{(1,2)}(\tau)
      & \equiv 
      -\frac{1}{2} \, \sum_{k=0}^\infty k \, 
      \chi_{20}^{(1,2)}(k) \, q^{ k^2 / 40} 
      \\
      & =
      q^{1/40} \,
      \sum_{k=0}^\infty (q)_k \,
      \sum_{j=0}^{k+1} q^{j^2} \,
      \begin{bmatrix}
        k+1 \\
        j
      \end{bmatrix}
      .
    \end{align*}
    The equalities in above formulae have same meaning
    with a case of trefoil~\cite{KHikami02c}.
    These are
    the Eichler integral of the Rogers--Ramanujan
    $q$-series,
    which is  the character of
    the Lee--Yang theory $\mathcal{M}(2,5)$.

  \item Knot    $T(3,4)$,
    \begin{align*}
      \widetilde{\Phi}^{(1,1)}(\tau)
      & \equiv 
      -\frac{1}{2} \, \sum_{k=0}^\infty k \, 
      \chi_{24}^{(1,1)}(k) \, q^{ k^2 / 48} 
      \\
      & =
      q^{1/48} \,
      \sum_{k=0}^\infty (q)_k
      \,
      \left(
        \sum_{j=0}^{\lfloor k/2 \rfloor} q^{2 j^2} \, 
        \begin{bmatrix}
          k \\
          2 \, j
        \end{bmatrix}
        +
        \sum_{j=0}^{\lfloor (k+1)/2 \rfloor} q^{2 j^2} \, 
        \begin{bmatrix}
          k+1 \\
          2 \, j
        \end{bmatrix}
      \right) ,
      \displaybreak[0]
      \\[2mm]
      \widetilde{\Phi}^{(1,2)}(\tau)
      & \equiv 
      -\frac{1}{2} \, \sum_{k=0}^\infty k \, 
      \chi_{24}^{(1,2)}(k) \, q^{ k^2 / 48} 
      \\
      & =
      2 \, q^{1/12} \,
      \sum_{k=0}^\infty
      (q^2; q^2)_k ,
      \displaybreak[0]
      \\[2mm]
      \widetilde{\Phi}^{(1,3)}(\tau)
      & \equiv 
      -\frac{1}{2} \, \sum_{k=0}^\infty k \, 
      \chi_{24}^{(1,3)}(k) \, q^{ k^2 / 48} 
      \\
      & =
      q^{25/48} \,
      \sum_{k=0}^\infty (q)_k
      \,
      \left(
        \sum_{j=0}^{\lfloor (k-1)/2 \rfloor} q^{2 j (j+1)} \, 
        \begin{bmatrix}
          k \\
          2 \, j +1
        \end{bmatrix}
        +
        \sum_{j=0}^{\lfloor k /2 \rfloor} q^{2 j (j+1)} \, 
        \begin{bmatrix}
          k+1 \\
          2 \, j +1
        \end{bmatrix}
      \right) ,
    \end{align*}
    These are the Eichler integral of the Slater's
    $q$-series~\cite{KHikami03b},
    which  is   the character of
    the Ising model $\mathcal{M}(3,4)$.

\end{itemize}
See that
infinite sums in 
all those expressions reduce to a finite sum in a case
$q \to \mathrm{e}^{2 \pi \mathrm{i}/N}$.

Asymptotic behavior of Kashaev's invariant,
\begin{equation*}
  \lim_{N\to \infty} \frac{2 \, \pi}{N} \,
  \log
  \langle \mathcal{K} \rangle_N ,
\end{equation*}
is conjectured~\cite{Kasha96b,MuraMura99a}
to give the
hyperbolic volume of the knot complement
$M= S^3 \setminus \mathcal{K}$.
In our case, the torus knot is not hyperbolic.
We can rather  expect
from  eqs.~\eqref{nearly_modular_Phi} and~\eqref{invariant_character} that
a value
\begin{equation}
  \label{log_limit_T}
  -\frac{(n  t - m  s)^2}{ s t} \, \pi^2 
  =
  - 4 \, \pi^2 \,
  \left(
    \Delta_{n,m}^{s,t} - \frac{c(s,t)-1}{24}
  \right) ,
\end{equation}
is related to the SU(2)
Chern--Simons invariant,
\begin{equation*}
  \CS(M)
  =
  \frac{1}{4} \int\limits_{M} \Tr
  \left(
    A \land \mathrm{d} A +
    \frac{2}{3} \, A \land A \land A
  \right) .
\end{equation*}
To see this fact,
we recall that
the fundamental group of $M=S^3 \setminus T(s,t)$ has a presentation
\begin{equation}
  \label{group_torus}
  \pi_1(M)
  =
  \langle x , y | x^s = y^t \rangle .
\end{equation}
As was shown in  Ref.~\citen{KirkKlas90a},
the Chern--Simons invariant from 
two  SU(2) representation $\rho_0$ and
$\rho_1$ of $\pi_1(M)$ satisfies
\begin{equation}
  \label{CS_difference}
  \CS(M ; \rho_1)
  - \CS(M ; \rho_0)
  =
  - 4 \, \pi^2 \,
  \int_0^1 \beta(z) \, \alpha^\prime(z) \,
  \mathrm{d} z .
\end{equation}
Here  $\alpha(z)$ and $\beta(z)$ are from the representation
$\rho_z$, $z \in [0,1]$,
of the
meridian $\mu$ and the longitude $\lambda$ up to conjugation,
\begin{align*}
  \rho_z(\mu)
  & =
  \begin{pmatrix}
    \mathrm{e}^{2 \pi \mathrm{i} \alpha(z)} &
    \\
    & \mathrm{e}^{-2 \pi \mathrm{i} \alpha(z)} 
  \end{pmatrix},
  &
  \rho_z(\lambda)
  & =
  \begin{pmatrix}
    \mathrm{e}^{2 \pi \mathrm{i} \beta(z)} &
    \\
    & \mathrm{e}^{-2 \pi \mathrm{i} \beta(z)} 
  \end{pmatrix} .
\end{align*}
In a case of complement~\eqref{group_torus} of  the torus knot,
the longitude $\lambda$ and the meridian $\mu$   are respectively given by
$x^s$ and $x^a \, y^b$, where
$a, b \in \mathbb{Z}$
satisfies
$a \, s + b \,t =1$.
As the longitude
$\lambda=x^s=y^t$ is a center of group, it is sent to $\pm 1$.
{}From relations
$ (x^a)^s = (x^s)^a$ and $(y^b)^t = (x^s)^b$
we see that $x^a$ and $y^b$ is conjugate to
\begin{align*}
  \rho(x^a)
  & \to 
  \begin{pmatrix}
    \mathrm{e}^{\pi \mathrm{i} n/s} &
    \\
    & \mathrm{e}^{- \pi \mathrm{i} n/s} 
  \end{pmatrix}  ,
  &
  \rho(y^b)
  & \to 
  \begin{pmatrix}
    \mathrm{e}^{\pi \mathrm{i} m/t} &
    \\
    & \mathrm{e}^{- \pi \mathrm{i} m/t} 
  \end{pmatrix}  ,
\end{align*}
where $n, m$ are integers.
Correspondingly we find that  a path of representation from a trivial
representation $z=0$ is given by
\begin{align*}
  \alpha(z)
  & =
  \frac{1}{2} \,
  \left(
    \frac{n}{s} + \frac{m}{t}
  \right)  \, z ,
  &
  \beta(z)
  & =
  \frac{s \, t}{2} \,
  \left(
    \frac{n}{s} + \frac{m}{t}
  \right)  .
\end{align*}
Here $\beta(z)$ is constant along this path representation since the
longitude is fixed to be $\pm 1$.
Substituting into eq.~\eqref{CS_difference}, we get
a quantity~\eqref{log_limit_T} as the Chern--Simons invariant of $M$ modulo
$2 \, \pi^2$.

\section{Concluding Remarks}

We have revealed intriguing  properties of the character of the minimal model
$\mathcal{M}(s,t)$.
We have shown that
Kashaev's invariant, \emph{i.e.}, a specific value of  the $N$-colored
Jones polynomial,
for the torus knot
$T(s,t)$ is regarded as the Eichler integral of the character for
$(n,m)=(s-1,1)$ with $q$ being the $N$-th root of unity.
It is natural to expect that general $(n,m)$ case is also related to
the quantum invariant of the 3-manifold.

As was shown in Ref.~\citen{KHikami03a}, the Eichler integral of the
affine $\widehat{\mathrm{su}}(2)_{m+2}$ character,
which is modular covariant with weight $3/2$,
gives  Kashaev's
invariant for
torus link $T(2, 2\, m)$ when $q$ is the $N$-th primitive root of unity.
As the torus knot and link are not hyperbolic,
we may regard
the hyperbolic manifold
as a (massive) deformation  of  the
conformal field
theory.

\section*{Acknowledgment}
The authors would like to thank
to H. Murakami for useful comments on early version of manuscript.
The work of KH is supported in part by the Sumitomo Foundation,
and
Grant-in-Aid for Young Scientists
from the Ministry of Education, Culture, Sports, Science and
Technology of Japan.

\bibliographystyle{gyoseki}

\end{document}